\documentclass[twocolumn]{aastex631}
\usepackage{lineno}
\usepackage{bm}

\begin{document}

\title{JWST Reveals a Surprisingly High Fraction of Galaxies Being Spiral-like at $0.5\leq z\leq4$}

\author{Vicki Kuhn}
\affiliation{Department of Physics and Astronomy, University of Missouri, Columbia, MO 65211, USA}

\author{Yicheng Guo}
\affiliation{Department of Physics and Astronomy, University of Missouri, Columbia, MO 65211, USA}

\author{Alec Martin}
\affiliation{Department of Physics and Astronomy, University of Missouri, Columbia, MO 65211, USA}

\author{Julianna Bayless}
\affiliation{Department of Physics and Astronomy, University of Missouri, Columbia, MO 65211, USA}

\author{Ellie Gates}
\affiliation{Department of Physics and Astronomy, University of Missouri, Columbia, MO 65211, USA}

\author{AJ Puleo}
\affiliation{Department of Physics and Astronomy, University of Missouri, Columbia, MO 65211, USA}

\correspondingauthor{Vicki Kuhn, Yicheng Guo}
\email{vkq9n@missouri.edu, guoyic@missouri.edu}



\begin{abstract}

Spiral arms are one of the most important features used to classify the morphology of local galaxies. The cosmic epoch when spiral arms first appeared contains essential clues to the evolution of disk galaxies. In this letter, we used James Webb Space Telescope (JWST) images from the Cosmic Evolution Early Release Science Survey to visually identify spiral galaxies with redshift $0.5\leq z\leq4$ and stellar mass $\geq10^{10}\; M_\odot$. Out of 873 galaxies, 216 were found to have a spiral structure. The spiral galaxies in our sample have higher star formation rates (SFRs) and larger sizes than non-spiral galaxies. We found the observed spiral fraction decreases from 48\% at $z\sim0.75$ to 8\% at $z\sim2.75$. These fractions are higher than the fractions observed with the Hubble Space Telescope (HST). We even detect possible spiral-like features at redshifts $z>3$. We artificially redshifted low redshift galaxies to high redshifts and re-inspected them to evaluate observational effects. By varying the input spiral fraction of the redshifted sample, we found that the input fraction of $\sim35$\% matches the observed fraction at $z=2-3$ the best. We are able to rule out spiral fractions being $<20$\% (2$\sigma$) and $<10$\% (3$\sigma$) for real galaxies at $z\sim3$. This fraction is surprisingly high and implies that the formation of spiral arms, as well as disks, was earlier in the universe.


\end{abstract}

\section{Introduction} \label{sec:intro}

Spiral structures are present in a majority of the galaxies that we see at low redshift \citep{sellwood_2022} and are the hubs of star formation \citep[e.g.,][]{Delgado_2010, Elmegreen_2011, grosbol/dottori_2012}. Understanding how and when spiral galaxies first existed in the universe has been a popular topic since their first discovery. The arms of spiral galaxies can range from grand design (having well-defined arms) to multi-arms to flocculent (patchy and discontinuous arms). Different types of spiral arms are believed to have different formation histories \citep{Elmegreen_1990}.

Grand design spirals can be explained by the density wave theory \citep{Lin_1964}. The theory states that the spiral arms are density waves that travel through the disk. The density waves triggers star formation downstream of the waves and induces a color jump in the direction of rotation inside the corotation radius \citep{grosbol_1998,gittins_2004,yu_2018}. Another theory is the formation of arms by swing amplification, which amplifies gravitational instabilities and can produce grand design and flocculent galaxies \citep{goldreich_1965, julian/toomre_1966}. Other methods of spiral formation include tidal effects from interactions with another galaxy \citep{byrd_1992}, bar instabilities \citep{kormendy_1979, dobbs_2010}, and the manifold theory \citep{athanassoula_2012}.

It is not quite clear what impact the number of spiral arms have on star formation. \cite{Hart_2017} found that the number of arms has little impact on the star formation rate, but \cite{Porter_Temple_2022} found that galaxies with more spiral arms tend to have higher star formation rates. The strength of spiral arms has been found to enhance star formation (e.g., \cite{kendall_2015,yu_2021}). Although spiral galaxies are well established by $z=1$, when they first appeared in the universe is still uncertain \citep{Conselice_2005, Elmegreen_2014}. HST-based studies have shown that galaxies around redshift $z=1$ have similar morphologies to galaxies at low redshifts and fit into the Hubble Sequence, but as redshift increases, galaxies tend to appear more irregular and clumpy \citep[e.g.,][]{Elmegreen_2005, elmegreen_2009, guo_2012, guo_2015, guo_2018, martin_2023}.

\cite{Margalef-Bentabol_2022} inspected the spiral fraction at high redshifts using HST imaging and found, after taking into account redshift effects, a spiral fraction of $\sim10$\% at $z=3$. However, the reddest HST filter, F160W, is only able to probe the rest-frame visible light up to $z\sim 2.8$. HST does not have a high enough sensitivity to resolve high redshift objects and the number of disk galaxies falls rapidly up to $z=3$ \citep{margalef-bentabol_2016}. Though \cite{Buitrago_2013} found that among massive galaxies at $z>3$, disks were reported as the higher morphology type. Indeed, at redshifts higher than $z=1$, spiral galaxies were thought to be sparse and very few have been found at $z>2$ with HST \citep[e.g.,][]{Law_2012}.


JWST is able to provide high sensitivity and resolution in the infrared and can probe rest frame visible light up to $z\sim 8$.  Already JWST has been making progress in updating our knowledge of the Hubble sequence in the early universe. Recent studies using JWST have shown that disk galaxies are prevalent up to redshifts of $z\sim6$ \citep{Ferreira_2022, Ferreira_2023, Kartaltepe_2023}. Disk galaxies tend to have spiral arms in the Local Universe and JWST’s improved sensitivity compared to HST can possibly detect spiral structures in high redshifted disk galaxies that HST was unable to provide. A couple of spiral galaxies were recently found at redshifts of $z=2.467$ \citep{Huang_2023} and $z=3.059$ \citep{Wu_2023} with the Atacama Large Millimeter/submillimeter Array (ALMA) and JWST and one barred spiral galaxy was discovered with JWST at $z\simeq3$ \cite{constantin_2023}.

In this letter, we explore the spiral fraction (the ratio of the number of spiral galaxies to the total number of galaxies in a redshift bin) up to $z\leq4$. We use JWST images to visually inspect spiral galaxies. We calculate the observed spiral fraction as well as correct for observational effects to derive an intrinsic spiral fraction. Throughout this letter we assume $\Omega_M=0.3$, $\Omega_\Lambda=0.7$, and $H_0=70$ km s$^{-1}$ Mpc$^{-1}$. AB magnitudes and a Chabrier initial mass function (IMF) are used throughout.

\section{Data and Sample} \label{sec:data}

We used the NIRCam imaging data of the Cosmic Evolution Early Release Science (CEERS) Survey \citep{Finkelstein_2023}. CEERS covered 100 square arcmin. of the Extended Groth Strip (EGS) during Cycle 1. The data was taken in ten separate pointings with seven filters: F115W, F150W, F200W, F277W, F356W, F410M, and F444W. We used images from the publicly released data from all ten pointings\footnote{\url{https://ceers.github.io}}. A description of the reduction and data release can be found in \cite{Bagley_2023} and \cite{Finkelstein_2023}, respectively.

Our sample of galaxies was selected from the footprint of CEERS NIRCam pointings and have been previously analyzed in the EGS catalog \citep{Stefanon_2017} from the Cosmic Assembly Near-IR Deep Extragalactic Legacy Survey (CANDELS) \citep{grogin_2011, koekemoer_2011}. This catalog includes photometric redshifts and stellar masses calculated through multiple SED-fitting codes and templates described in \cite{dahlen_2013} and \cite{mobasher_2015}, respectively. We use the median values from these calculations to construct the bounds of our sample. We excluded objects that fell between the gaps in the NIRCam detectors. We then chose objects with stellar masses greater than $10^{10}$ M$_\odot$ and were within the redshift range $0.5\leq z\leq4$. We excluded objects with a CLASS\_STAR$>$0.8 to remove likely star candidates. With these parameters, we have a sample of 873 galaxies.

\section{Methods} \label{sec:methods}

For each galaxy, we made a 100$\times$100 pixel ($3''\times3''$) cutout in its rest frame V band, which corresponds to F115W  at $z=0.5-1.36$, F150W at $z=1.36-2.09$, F200W at $z=2.09-3.18$, or F277W at $z=3.18-4$. The cutout was displayed as an image with a logarithmic gray scale. All six co-authors of this paper visually inspected each image and voted ``yes" or ``no" for whether they observed spiral structure.

We utilized two methods for calculating the spiral fraction. First, the binary decision identifies spiral galaxies if they have been given a ``yes" vote by at least three people (i.e. $\geq50\%$). Those with less than three ''yes" votes are considered non-spirals. With this decision, the spiral fraction in a given redshift bin is the ratio of the number of spiral galaxies to the number of all galaxies in that bin. Second, the probability decision calculates the likelihood of a galaxy being spiral by taking the ratio of the number of ``yes" votes to the number of the total votes. In this method, each galaxy has a spiral probability ranging between 0 and 1. The probability decision fraction in a given redshift bin is then the average spiral probability of all galaxies in that bin.

\begin{figure*}[ht!]
\epsscale{0.8}
\plotone{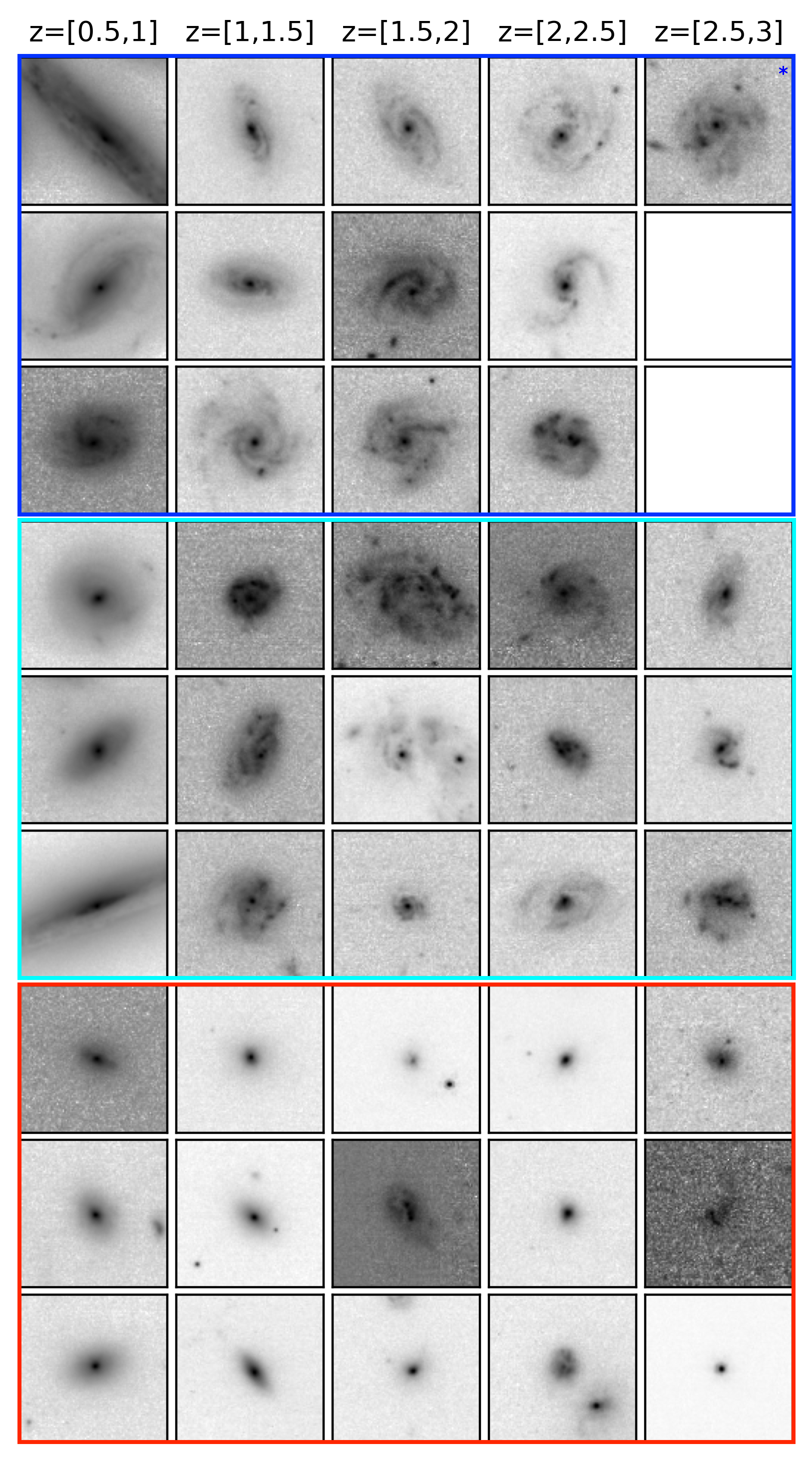}
\caption{Examples of galaxies in our sample. Redshifts increase from left to right as indicated by the labels. Galaxies in the top three rows (outlined in the large blue box) were classified as spirals by all inspectors (except for the top right corner with an asterisk which only had 5 inspectors vote spiral). The two blank cutouts signify no other spiral galaxies identified in that redshift bin. The middle three rows (outlined in the large cyan box) show galaxies with spiral votes of 3-4. The bottom three rows (outlined in the large red box) show non-spiral galaxies.
\label{fig:fig1}}
\end{figure*}

\section{Results} \label{sec:results}

\subsection{Observed Spiral Fraction} \label{subsec:fraction}

\begin{figure*}[ht!]
\plotone{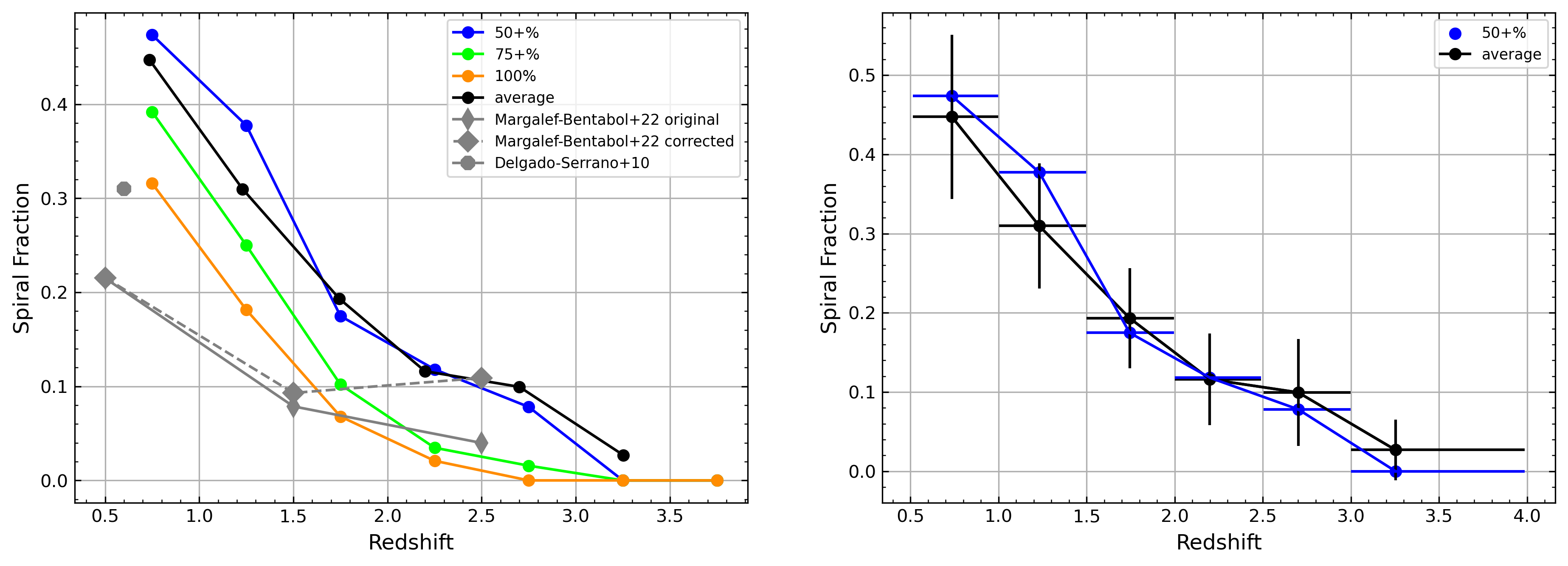}
\caption{Fraction of spiral galaxies as a function of redshift. The blue points represent the galaxies in our sample that had at least 3 voters classify it as a spiral, the lime represents 5+ votes, and the orange all 6 voters classified as spiral (binary decision). The black points represent the average score of our sample (probability decision). The gray hexagon represents the fraction found from \cite{Delgado_2010}. The thin and thick diamonds show the uncorrected and corrected fractions after taking into account redshift effects from \cite{Margalef-Bentabol_2022}, respectively. The error bars in the right panel shows 3$\sigma$ standard error in the y-axis and the 16th and 84th percentile values in the x-axis.
\label{fig:fig2}}
\end{figure*}

Out of the 873 galaxies in the sample, 216 were classified as spiral using the binary decision with 108 receiving a ``yes" vote from all classifiers. In Figure \ref{fig:fig1}, we present a sample of galaxies in our data set. The columns show increasing redshift bins, from left to right, while rows from top to bottom show the range of galaxies classified as spiral to non-spiral. Spiral structure is easier to see at the lower redshift ranges and becomes less pronounced at higher redshifts.

We show the evolution of the spiral fraction as a function of redshift in Figure \ref{fig:fig2}. The binary spiral fraction (blue line) decreases starting from 48\% at $z=0.5-1$ to 15\% at $z=2-2.5$ then down to 0\% at $z=3-3.5$. We also calculate a stricter binary decision fraction at 75\% (5+ ``yes" votes, lime line) and 100\% (6 ``yes" votes, orange line). These results show a lower spiral fraction compared to the 50\% binary decision fraction at all redshifts. The probability spiral fraction (black line) is 45\% at $z=0.5-1$ and drops to 3\% at $z=3-3.5$. These two decisions are consistent with each other (see right panel of Fig. \ref{fig:fig2}). Later, we will choose the decision method that best illustrates our results. We observe a steep drop in the spiral fraction before $z=2$, though overall remains consistent with the fractions at the right end of the plot.


We compare our results to \cite{Margalef-Bentabol_2022}. They used data from the fourth release of Galaxy Zoo which showed RGB images from HST in the COSMOS, GOODS-S, and UDS fields. Their selection criteria had $\geq10$ classifiers per galaxy and a cutoff of $>50$\% for spiral galaxies. Our spiral fraction is significantly higher, more than double their spiral+clumpy spiral fraction (thin gray diamonds), at any given redshift below $z=3$. This result is not surprising given the resolution and sensitivity of HST compared with JWST. When \cite{Margalef-Bentabol_2022} took into account the redshift related observational effects of their sample (large gray diamonds), their spiral fraction at $z=2.5$ is comparable to ours. However, we have not yet corrected for redshift effects.

\cite{Delgado_2010} conducted a study comparing local galaxies to distant galaxies ($0.4\leq z \leq 0.8$) and found that about 31\% of the distant galaxies are spirals. They defined spiral galaxies by if the bulge was redder than the disk, arms were visually seen and the disk was symmetric, and the bulge-to-total ratio is smaller than 0.5. Their spiral fraction is also lower than our results from both the binary and probability decisions, implying a higher resolution from JWST.

\subsection{Star Formation and Size} \label{subsec:starformation}

\begin{figure*}
\plotone{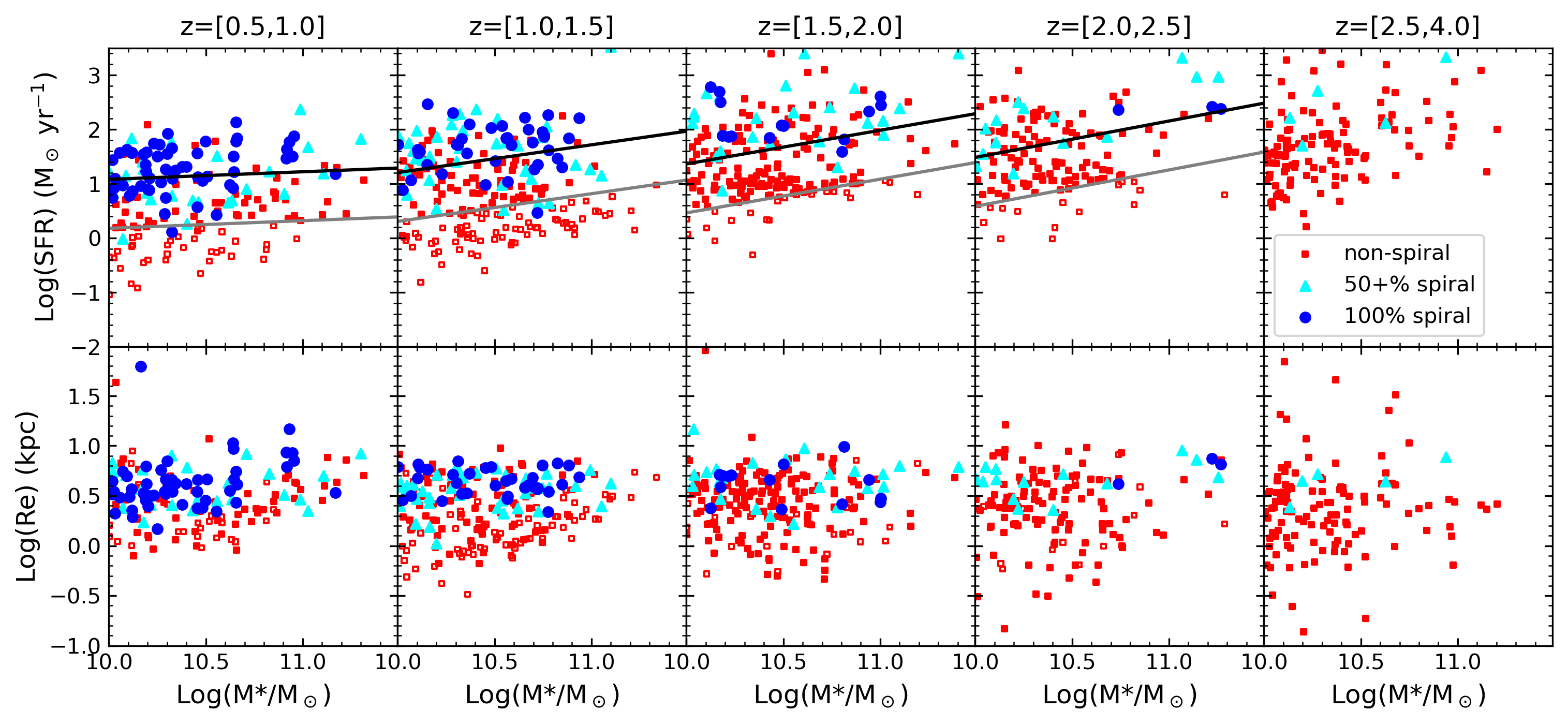}
\caption{The top row shows the SFR as a function of stellar mass in separate redshift bins. Blue points show spiral galaxies with 6 ``yes" votes, cyan triangles show spiral galaxies with 3-5 ``yes" votes, and red squares show non-spiral galaxies, 2 or less ``yes" votes. Non-spiral galaxies below the star-forming region are shown as open red squares. The black line shows the star-forming main sequence line \citep{whitaker_2014} and the gray line (0.9 dex below the black line) shows the lower boundary of the star-forming region. The bottom row shows the half-light radius as a function of stellar mass.
\label{fig:fig5}}
\end{figure*}

In Figure \ref{fig:fig5}, we show the SFR and half-light radius as a function of stellar mass in each of our redshift bins. We use the $SFR^{UV}_{corr}$ values from \cite{Barro_2019}, which were derived from UV luminosities using the relation from \cite{Kennicutt_1998} and corrected for dust using the slope of the UV continuum emission to the ratio of the UV to IR luminosities. The majority of our galaxies classified as having a spiral structure (blue points and cyan triangles) lie above or in the star-forming region (gray line). As redshift increases, more spiral galaxies are seen above the main sequence (black line). Spiral galaxies in our sample with weak arms may be misidentified as non-spiral, especially at higher redshifts, due to observational effects. \cite{yu_2021}, using a sample of low redshift spiral galaxies ($z\leq0.03$), found that spiral galaxies above the star-forming main sequence are stronger than those below it, further highlighting that some spiral galaxies may have been missed which would increase the number of spiral galaxies found at higher redshifts.

The bottom panel of Fig. \ref{fig:fig5} shows that spiral galaxies have larger sizes than non-spiral galaxies. Some small size galaxies that were classified as non-spirals may in fact be spiral galaxies but were missed during the visual inspection due to noise or PSF effects. We also found there is little change in size, for spiral galaxies, as redshift increases. The lack of changes in size may be due to sample selection (missing disk galaxies) or bandpass effects since we used sizes from HST's F160W filter \citep{vanderwel_2012}.


\subsection{Intrinsic Spiral Fraction}
\label{subsec:artificial}

Due to the cosmological dimming and the sensitivity of CEERS observations, spiral galaxies can be hard to identify at high redshifts. This issue causes observed spiral fractions to be lower than the real fraction (see \cite{Margalef-Bentabol_2022}). Artificially shifting galaxies is a useful method to understand the biases and uncertainties in measurements between low and high-redshift galaxies (see \cite{giavalisco_1996, barden_2008, yu_2023}). To correct for this issue, we selected a representative sample of galaxies between redshifts 0.5 and 0.8 to artificially shift to higher redshifts. This sample, called the unredshifted sample, contained 101 galaxies and had sizes between $0.1''$ and $2''$. Our procedure is outlined below. We use the target redshift bin $2<z<2.5$ as an example.
\begin{enumerate}[noitemsep]
    \item We took the rest frame V-band of the unredshifted sample (F115W) and calculated its median flux ($f_{unz}$). We then calculated the median flux of the real galaxies at $2<z<2.5$ in their rest frame V-band (F200W), $f_z$. Then, we scaled down the flux of each galaxy in the unredshifted sample by a factor of $f_z/f_{unz}$.
    \item The flux-down images were then downsized to take into account (a) the intrinsic size evolution of disk galaxies and (b) the angular diameter distance ($D_A$). For the former, we use the relation of $(1+z)^{-0.75}$ from \cite{van_der_wel_2014} to determine a factor to shrink the image size at high redshift. For the latter, we use the ratio of $D_A(z_{high})/D_A(z=0.75)$ as the second factor. The flux-down images were shrunk by the product of the two factors.
    \item The downsized images were then convolved with a Gaussian kernel, to take into account the PSF difference betwen F115W and F200W. The FWHM of the kernel was calculated by $\sqrt{F200W FWHM^2-F115W FWHM^2}$, where the $F115W FWHM^2$ is the PSF in F115W downsized according to Step 2 and $F200W FWHM^2$ corresponds to the redshift bin we are using. 
    \item We found a blank sky region in F200W and added the convolved images to these regions.
\end{enumerate}
We repeated this procedure for all of our redshift bins. Each redshifted image was visually inspected by each of the six co-authors.


\begin{figure*}[ht!]
\plotone{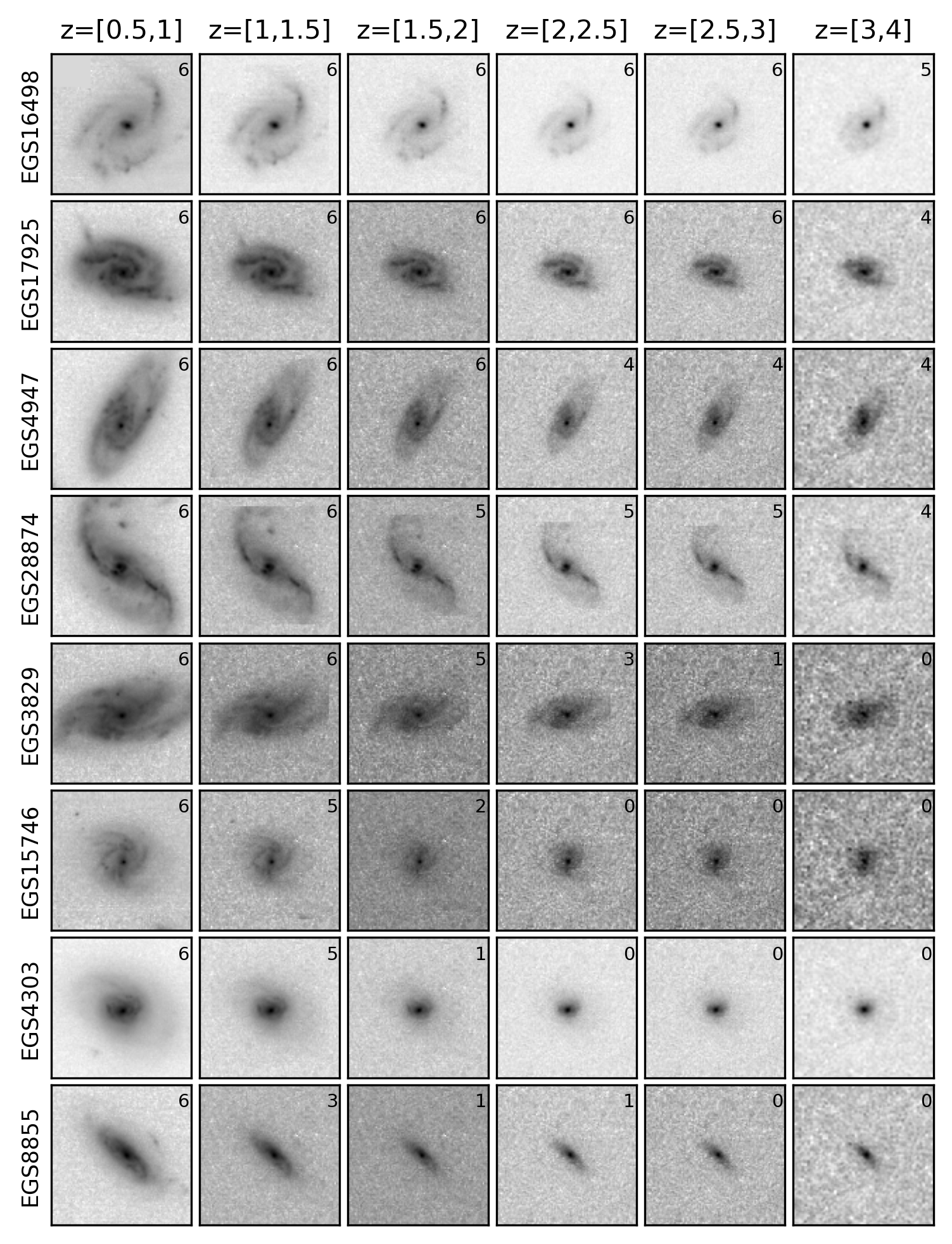}
\caption{Examples of galaxies within our artificially redshifted sample. Each row shows one galaxy at differing redshifts. The number of classifiers that designated the galaxy as spiral is indicated in the top right corner of each image. The top four rows show examples of galaxies with a clear spiral structure across all the redshift bins. The bottom four show galaxies whose spiral structure is not detectable at higher redshifts.
\label{fig:fig3}}
\end{figure*}

Figure \ref{fig:fig3} shows some example galaxies from this sample set in bins of increasing redshift. Top four rows show examples of redshifted spirals that have clear spiral features out to higher redshifts and the bottom four rows show examples of galaxies where the classifiers were unable to observe spiral features at higher redshifts.

\begin{figure*}[t]
\plotone{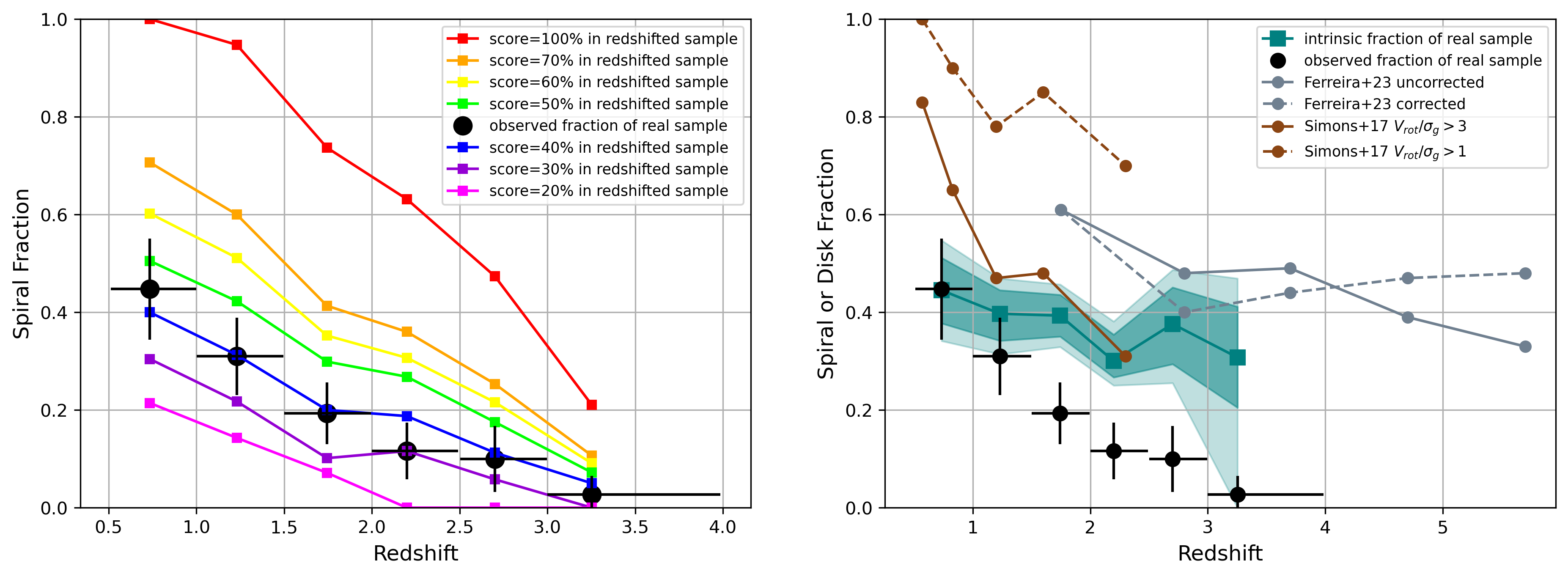}
\caption{The left panel shows the fraction of spiral galaxies as a function of redshift from our subsample with varying intrinsic spiral fractions. The big black points show the results of our probability decision. The x and y errors are the median of the redshift in each bin and 3$\sigma$ standard error, respectively. Each of the different colored lines shows how the observed spiral fraction of a sample with different intrinsic spiral fraction at $z=0.5-0.8$ would change when the sample was shifted to different redshifts. The intrinsic fraction that matches the observed spiral fraction at a given redshift would be considered the true spiral fraction. The right panel shows the intrinsic fraction of the real sample (teal squares) across different redshifts compared with the fraction of disk galaxies (brown and gray lines) that have been reported in the literature.
\label{fig:fig4}}
\end{figure*}

To account for the redshift effects, instead of calculating a correction factor to scale up our observed spiral fraction, we used a more ``forward modeling" method (left panel of Figure \ref{fig:fig4}). We constructed sub-samples from the unredshifted sample with various spiral fractions ranging from 100\% probability (shown in red) to 20\% probability (shown in pink) by removing spiral or non-spiral galaxies. We call these fractions intrinsic fractions. We then calculated the redshifted spiral fraction for each sub-sample. The intrinsic spiral fraction that produces a redshifted spiral fraction matching our observed spiral fraction (black points) would be considered as the true spiral fraction in the universe at that redshift. The intrinsic fraction and associated errors are calculated using a linear interpolation. For example, at $2.5<z<3$, the observed fraction falls between an intrinsic fraction of 40\% (blue line) and 30\% (purple line). We then use linear interpolation to determine that the best matched intrinsic spiral fraction is 37.6\%, which is shown with a teal square at $2.5<z<3$ in the right panel of Fig. \ref{fig:fig4}. Similarly, we used linear interpolation on the error bars of the observed fraction to determine the upper (48.8\%) and lower (25.5\%) limits with 3$\sigma$.

Overall, the right panel of Figure \ref{fig:fig4} shows the intrinsic spiral fraction of 40\% (blue line) matches our observed spiral fraction very well, implying a high spiral fraction at high redshifts. The intrinsic fraction (teal line) remains mostly flat until $z\sim2.75$, then slightly declines at $z>3$. The light (dark) teal shaded area shows the 3$\sigma$ (2$\sigma$) standard error of the intrinsic fraction. Within 3$\sigma$ of the intrinsic values, the lower limit of spiral fraction is 33\% at $z\sim1.75$, 25\% at $z\sim2.25$, and 25\% at $z\sim2.75$. These limits are higher than that previously reported in the literature (e.g., \cite{margalef-bentabol_2016}). At $z>3$, observational effects are so severe that the sub-samples with the intrinsic spiral fraction $\leq40\%$ all approach a close-to-zero observed spiral fraction. The observed fraction loses the power of constraining the 3$\sigma$ lower limit; but within 2$\sigma$, our lower limit is still high (20\%).




\section{Discussion} \label{sec:discussion}

\subsection{Spiral and Disk Formation} \label{subsec:diskformation}

Our results of a high spiral fraction at $z\gtrsim2$ (both the median and the lower limit) indicate an early disk and spiral arm formation in the universe. Recent JWST results on galaxy morphology have found that about 40-50\% of galaxies at $z = 3-6$ show a disk morphology \citep[e.g.,][]{Ferreira_2022, Ferreira_2023, Kartaltepe_2023}. Some rotating cold gas disks have already been reported up to redshift $z\sim4.75$ by JWST and ALMA \citep[e.g.,][]{rizzo_2020, lelli_2021, fraternali_2021}. Our results, using spiral arms as an indicator of rotating stellar disk, are consistent with this scenario. In the right panel of Figure \ref{fig:fig4}, we compare the disk fraction from \cite{Ferreira_2023} (dashed gray line) to our spiral fraction. Both fractions broadly match at $z\sim3$, implying that stellar disks, either identified through the overall morphology of galaxies or through the existence of spiral arms, started to dominate a large fraction of galaxies when the age of the universe is around 2 Gyrs ($z\sim3$). 

To further understand the condition of spiral formation, we also compare our spiral fraction to the fraction of settled gas disks in the right panel of Figure \ref{fig:fig4}. \cite{simons_2017} reported the disk fraction of star-forming galaxies at $0.5 \lesssim z \lesssim 2.5$. They identified disk galaxies as those with the gas rotation velocity-to-velocity dispersion ratio, $V_{rot}/\sigma_g$, were greater than 3 (solid brown line) or 1 (dashed brown line). Our spiral fraction (the teal region) is lower than the fraction of galaxies with $V_{rot}/\sigma_g > 1$, but matches that of galaxies with $V_{rot}/\sigma_g > 3$ remarkably well. This agreement implies that both spiral formation and disk settling occurred at the same cosmic time. Putting all above pieces together, our results support such a disk and spiral formation scenario: disk galaxies appear early in the universe ($z\gtrsim4-5$) with dynamically hot gaseous disk ($V_{rot}/\sigma_g \lesssim 3$, e.g., \cite{tsukui_2021,nelson_2023}). These disks settled down to dynamically cold ($V_{rot}/\sigma_g > 3$) gaseous disks around $z\sim3-4$, and meanwhile dynamically cold stellar disks, indicated by the existence of spiral arms, also started to appear in a high frequency. 

Interestingly, our spiral fraction matches the gas kinematics measured from nebular emission around star-forming regions \citep[e.g.,][]{simons_2017}. Recent ALMA observations on the kinematics of CO and [CII] gas show a constantly high $V_{rot}/\sigma_g \sim 10$ from $z=0$ to $z=5$ \citep[e.g.,][]{lelli_2018, rizzo_2020, lelli_2021, rizzo_2021}. These results indicate that spiral formation and maintenance are affected by both warm gas (e.g., gas in both disk and in outflow traced by H$\alpha$ or other nebular emission lines) and the cold gas reservoir traced by CO or [CII]. When disks cool down, spirals form from either internal or external perturbations as enhanced star formation regions. The galactic spiral shocks can contribute a substantial amount of kinetic energy to heat up the cold gas \citep{kim_2010}. H$\alpha$ emission can also transfer energy from new stars to interstellar gas. These energy transfers are important to maintain spiral structures, because without them, spirals would gain heat due to scattering effects at the corotation \citep{binney_2008}, eventually leading to their fading and transformation into lenticular galaxies.

Our results also shed new light on the relation between spirals and other galactic substructures. \cite{Elmegreen_2005} and \cite{dekel_2009} showed that gas-rich disks at high redshift are highly turbulent and have giant star-forming clumps formed by gravitational instabilities. At later stages, hot stars increase the velocity dispersion of the disks (and hence the Toomre Q-parameter, \cite{toomre_1964} and \cite{genzel_2011}) to stabilize them. Meanwhile, merging bulges, possibly form mergers of clumps \citep{elmegreen_2008,ceverino_2010}, can also stabilize the disks. Therefore, gravitational instabilities tend to produce spirals rather than clumps \citep{bournaud_2009}. Based on HST observations, the spiral structure in the disks then occurs at $1.4<z<1.8$, when the disks transition from velocity dispersion dominated to rotation dominated \citep{Elmegreen_2014}. These pre-JWST results outlined an evolutionary scenario of galactic structures: starting from gas-rich, turbulent disk, followed by clumps, then hot stellar disk+bulge, and eventually to spirals. Our results, however, show a more mixed and earlier scenario, starting from gas-rich, turbulent disk to the co-existence of clumps and spirals at high redshifts. \cite{bland-hawthorn_2023} and 
\cite{bland-hawthorn_2024} shows that substructures (e.g., bars and spirals) emerge at early redshifts in gas-rich turbulent
disks. In fact, the majority of example spiral galaxies at $z>1.5$ in our Figure \ref{fig:fig1} show the co-existence of clumps and spirals. This scenario is also consistent with the high fraction of clumpy galaxies at $1.5<z<3.0$ \citep{guo_2015}.



\subsection{Possible Caveats} \label{subsec:caveats}

Our studies have a few possible caveats. First, galaxies in merger or interaction can be misclassified as spiral galaxies, and vice versa. In a late stage of merger, the long tidal tail of the system can be identified as spiral arms. This situation is more severe for galaxies at $z>2$, as the merger fraction is believed to be higher then \citep[e.g.,][]{duncan_2019}. An example is the galaxy on the top right corner of Figure \ref{fig:fig1}, which has both a tail (or an arm) and a disturbed disk. If these systems are indeed mergers rather than disks, they would result in an overestimate of spiral fraction at high redshifts. 

To evaluate the impact of mergers, we used the formula of \cite{wen_2016} to estimate the fraction of galaxies with long tidal tails in mergers. According to \cite{wen_2016}, this long tidal tail fraction among star-forming galaxies increases with redshift as $f_{tail}(z) = 0.64(1+z)^2$ to $z\sim1$. If we extrapolate their formula to $z\sim3$ (or $z\sim2$), about 10.2\% (or 5.8\%) of our sample would be long-tailed merging galaxies. We subtracted $f_{tail}(z)$ from our observed spiral fraction in the left panel of Figure \ref{fig:fig4} (black circles and error bars) and recalculated the true spiral fraction (teal squares) in the right panel of Figure \ref{fig:fig4}. After this subtraction, the true spiral fraction is almost not changed for $z\lesssim2$, but drops to about 20\% at $z\sim3$. This new fraction is still high compared to the literature, validating our general results. Moreover, this simple fraction-subtraction method may underestimate the true spiral fraction, because we did not subtract the same effect from our redshifted samples (i.e., the color curves in the left panel of Figure \ref{fig:fig4}). If we did so, the recovered true spiral fraction would be the same as before (i.e., $\sim$40\% at $z\sim3$), but the uncertainty range would be much larger, because after the subtraction, most redshifted samples have a close-to-zero fraction at $z\sim3$ so that the observation is not able to distinguish curves of different intrinsic fractions effectively.

The second possible caveat is the personal biases regarding visual inspection. To determine how significantly inspectors' biases affected our results, we repeated our spiral fraction measurements using bootstrapping on the inspectors. We randomly drew six inspectors with replacement in our inspector pool. We recalculated the spiral fraction based on the results of the redrawn inspectors. We then repeated this drawing 100 times to calculate the mean and standard deviation of the spiral fractions, for both the probability and binary decisions. We found that the results from this method were consistent with our previous results in the right panel of Fig. \ref{fig:fig2}. Therefore, we conclude that statistically, our results are not significantly affected by personal biases.

\section{Conclusion} \label{sec:conclusion}

We used the JWST CEERS images to visually inspect galaxies up to $z\sim4$. We used the binary decision method to classify spiral galaxies and then used this, and the probability decision, to calculate the spiral fraction at multiple redshifts. The observed spiral fraction decreases with increasing redshift, from $\sim43$\% at $z=1$ to $\sim4$\% at $z=3$. We found that the spiral galaxies in our sample have higher SFRs as well as sizes compared to non-spiral galaxies. We artificially redshifted low redshift galaxies to high redshifts to account for observational effects in order to recover the true spiral fraction at different redshifts. We found the intrinsic spiral fraction is about $\sim40$\% across all redshifts. We were able to set lower limits on the spiral fraction at redshifts 1.75 (33\%), 2.25 (25\%), 2.75 (25\%) within 3$\sigma$, and 3.25 (20\%) within 2$\sigma$. Our results of a high spiral fraction (and its lower limit) at $z\sim3$ supports the scenario of disk galaxies forming at $z\gtrsim4-5$ and settling down to dynamically cold, thin disks by $z\lesssim3$.




In the future, this study can be improved by larger and deeper observations (e.g. PRIMER, COSMOS-Web) to expand our sample for further analysis on the spiral fraction at high redshifts. Moreover, comparisons with simulations can set tighter constraints on the processes that govern formation of spiral features and disk galaxies.

\

We thank the anonymous referee for careful review and valuable comments that led to the improvement of this letter. We thank Gregory Rudnick and Haojing Yan for their valuable suggestions. This work is supported by the NASA-Missouri Space Grant Consortium No. 80NSSC20M0100 and the University of Missouri Research Council Grant URC-23-036. JB and EG thank the support of Colonel Arthur C. Allen Scholarship fund.

\vspace{5mm}

\bibliography{sample631}{}

\begin{thebibliography}{}
\expandafter\ifx\csname natexlab\endcsname\relax\def\natexlab#1{#1}\fi
\providecommand{\url}[1]{\href{#1}{#1}}
\providecommand{\dodoi}[1]{doi:~\href{http://doi.org/#1}{\nolinkurl{#1}}}
\providecommand{\doeprint}[1]{\href{http://ascl.net/#1}{\nolinkurl{http://ascl.net/#1}}}
\providecommand{\doarXiv}[1]{\href{https://arxiv.org/abs/#1}{\nolinkurl{https://arxiv.org/abs/#1}}}

\bibitem[{{Athanassoula}(2012)}]{athanassoula_2012}
{Athanassoula}, E. 2012, \mnras, 426, L46,
  \dodoi{10.1111/j.1745-3933.2012.01320.x}

\bibitem[{Bagley {et~al.}(2023)Bagley, Finkelstein, Koekemoer, Ferguson, Haro,
  Dickinson, Kartaltepe, Papovich, Pérez-González, Pirzkal, Somerville,
  Willmer, Yang, Yung, Fontana, Grazian, Grogin, Hirschmann, Kewley,
  Kirkpatrick, Kocevski, Lotz, Medrano, Morales, Pentericci, Ravindranath,
  Trump, Wilkins, Calabrò, Cooper, Costantin, de~la Vega, Hilbert, Hutchison,
  Larson, Lucas, McGrath, Ryan, Wang, \& Wuyts}]{Bagley_2023}
Bagley, M.~B., Finkelstein, S.~L., Koekemoer, A.~M., {et~al.} 2023, The
  Astrophysical Journal Letters, 946, L12, \dodoi{10.3847/2041-8213/acbb08}

\bibitem[{{Barden} {et~al.}(2008){Barden}, {Jahnke}, \&
  {H{\"a}u{\ss}ler}}]{barden_2008}
{Barden}, M., {Jahnke}, K., \& {H{\"a}u{\ss}ler}, B. 2008, \apjs, 175, 105,
  \dodoi{10.1086/524039}

\bibitem[{Barro {et~al.}(2019)Barro, Pérez-González, Cava, Brammer, Pandya,
  Moral, Esquej, Domínguez-Sánchez, Pampliega, Guo, Koekemoer, Trump, Ashby,
  Cardiel, Castellano, Conselice, Dickinson, Dolch, Donley, Briones, Faber,
  Fazio, Ferguson, Finkelstein, Fontana, Galametz, Gardner, Gawiser,
  Giavalisco, Grazian, Grogin, Hathi, Hemmati, Hernán-Caballero, Kocevski,
  Koo, Kodra, Lee, Lin, Lucas, Mobasher, McGrath, Nandra, Nayyeri, Newman,
  Pforr, Peth, Rafelski, Rodríguez-Munoz, Salvato, Stefanon, van~der Wel,
  Willner, Wiklind, \& Wuyts}]{Barro_2019}
Barro, G., Pérez-González, P.~G., Cava, A., {et~al.} 2019, The Astrophysical
  Journal Supplement Series, 243, 22, \dodoi{10.3847/1538-4365/ab23f2}

\bibitem[{{Binney} \& {Tremaine}(2008)}]{binney_2008}
{Binney}, J., \& {Tremaine}, S. 2008, {Galactic Dynamics: Second Edition}

\bibitem[{{Bland-Hawthorn} {et~al.}(2024){Bland-Hawthorn}, {Tepper-Garcia},
  {Agertz}, \& {Federrath}}]{bland-hawthorn_2024}
{Bland-Hawthorn}, J., {Tepper-Garcia}, T., {Agertz}, O., \& {Federrath}, C.
  2024, arXiv e-prints, arXiv:2402.06060, \dodoi{10.48550/arXiv.2402.06060}

\bibitem[{{Bland-Hawthorn} {et~al.}(2023){Bland-Hawthorn}, {Tepper-Garcia},
  {Agertz}, \& {Freeman}}]{bland-hawthorn_2023}
{Bland-Hawthorn}, J., {Tepper-Garcia}, T., {Agertz}, O., \& {Freeman}, K. 2023,
  \apj, 947, 80, \dodoi{10.3847/1538-4357/acc469}

\bibitem[{{Bournaud} {et~al.}(2009){Bournaud}, {Elmegreen}, \&
  {Martig}}]{bournaud_2009}
{Bournaud}, F., {Elmegreen}, B.~G., \& {Martig}, M. 2009, \apjl, 707, L1,
  \dodoi{10.1088/0004-637X/707/1/L1}

\bibitem[{{Buitrago} {et~al.}(2013){Buitrago}, {Trujillo}, {Conselice}, \&
  {H{\"a}u{\ss}ler}}]{Buitrago_2013}
{Buitrago}, F., {Trujillo}, I., {Conselice}, C.~J., \& {H{\"a}u{\ss}ler}, B.
  2013, \mnras, 428, 1460, \dodoi{10.1093/mnras/sts124}

\bibitem[{{Byrd} \& {Howard}(1992)}]{byrd_1992}
{Byrd}, G.~G., \& {Howard}, S. 1992, \aj, 103, 1089, \dodoi{10.1086/116128}

\bibitem[{{Ceverino} {et~al.}(2010){Ceverino}, {Dekel}, \&
  {Bournaud}}]{ceverino_2010}
{Ceverino}, D., {Dekel}, A., \& {Bournaud}, F. 2010, \mnras, 404, 2151,
  \dodoi{10.1111/j.1365-2966.2010.16433.x}

\bibitem[{Conselice {et~al.}(2005)Conselice, Blackburne, \&
  Papovich}]{Conselice_2005}
Conselice, C.~J., Blackburne, J.~A., \& Papovich, C. 2005, The Astrophysical
  Journal, 620, 564, \dodoi{10.1086/426102}

\bibitem[{{Costantin} {et~al.}(2023){Costantin}, {P{\'e}rez-Gonz{\'a}lez},
  {Guo}, {Buttitta}, {Jogee}, {Bagley}, {Barro}, {Kartaltepe}, {Koekemoer},
  {Cabello}, {Corsini}, {M{\'e}ndez-Abreu}, {de la Vega}, {Iyer}, {Bisigello},
  {Cheng}, {Morelli}, {Arrabal Haro}, {Buitrago}, {Cooper}, {Dekel},
  {Dickinson}, {Finkelstein}, {Giavalisco}, {Holwerda}, {Huertas-Company},
  {Lucas}, {Papovich}, {Pirzkal}, {Seill{\'e}}, {Vega-Ferrero}, {Wuyts}, \&
  {Yung}}]{constantin_2023}
{Costantin}, L., {P{\'e}rez-Gonz{\'a}lez}, P.~G., {Guo}, Y., {et~al.} 2023,
  \nat, 623, 499, \dodoi{10.1038/s41586-023-06636-x}

\bibitem[{{Dahlen} {et~al.}(2013){Dahlen}, {Mobasher}, {Faber}, {Ferguson},
  {Barro}, {Finkelstein}, {Finlator}, {Fontana}, {Gruetzbauch}, {Johnson},
  {Pforr}, {Salvato}, {Wiklind}, {Wuyts}, {Acquaviva}, {Dickinson}, {Guo},
  {Huang}, {Huang}, {Newman}, {Bell}, {Conselice}, {Galametz}, {Gawiser},
  {Giavalisco}, {Grogin}, {Hathi}, {Kocevski}, {Koekemoer}, {Koo}, {Lee},
  {McGrath}, {Papovich}, {Peth}, {Ryan}, {Somerville}, {Weiner}, \&
  {Wilson}}]{dahlen_2013}
{Dahlen}, T., {Mobasher}, B., {Faber}, S.~M., {et~al.} 2013, \apj, 775, 93,
  \dodoi{10.1088/0004-637X/775/2/93}

\bibitem[{{Dekel} {et~al.}(2009){Dekel}, {Sari}, \& {Ceverino}}]{dekel_2009}
{Dekel}, A., {Sari}, R., \& {Ceverino}, D. 2009, \apj, 703, 785,
  \dodoi{10.1088/0004-637X/703/1/785}

\bibitem[{{Delgado-Serrano, R.} {et~al.}(2010){Delgado-Serrano, R.}, {Hammer,
  F.}, {Yang, Y. B.}, {Puech, M.}, {Flores, H.}, \& {Rodrigues,
  M.}}]{Delgado_2010}
{Delgado-Serrano, R.}, {Hammer, F.}, {Yang, Y. B.}, {et~al.} 2010, A\&A, 509,
  A78, \dodoi{10.1051/0004-6361/200912704}

\bibitem[{Dobbs {et~al.}(2010)Dobbs, Theis, Pringle, \& Bate}]{dobbs_2010}
Dobbs, C.~L., Theis, C., Pringle, J.~E., \& Bate, M.~R. 2010, Monthly Notices
  of the Royal Astronomical Society, 403, 625,
  \dodoi{10.1111/j.1365-2966.2009.16161.x}

\bibitem[{{Duncan} {et~al.}(2019){Duncan}, {Conselice}, {Mundy}, {Bell},
  {Donley}, {Galametz}, {Guo}, {Grogin}, {Hathi}, {Kartaltepe}, {Kocevski},
  {Koekemoer}, {P{\'e}rez-Gonz{\'a}lez}, {Mantha}, {Snyder}, \&
  {Stefanon}}]{duncan_2019}
{Duncan}, K., {Conselice}, C.~J., {Mundy}, C., {et~al.} 2019, \apj, 876, 110,
  \dodoi{10.3847/1538-4357/ab148a}

\bibitem[{{Elmegreen}(1990)}]{Elmegreen_1990}
{Elmegreen}, B.~G. 1990, Annals of the New York Academy of Sciences, 596, 40,
  \dodoi{10.1111/j.1749-6632.1990.tb27410.x}

\bibitem[{{Elmegreen}(2011)}]{Elmegreen_2011}
{Elmegreen}, B.~G. 2011, in EAS Publications Series, Vol.~51, EAS Publications
  Series, ed. C.~{Charbonnel} \& T.~{Montmerle}, 19--30,
  \dodoi{10.1051/eas/1151002}

\bibitem[{{Elmegreen} {et~al.}(2008){Elmegreen}, {Bournaud}, \&
  {Elmegreen}}]{elmegreen_2008}
{Elmegreen}, B.~G., {Bournaud}, F., \& {Elmegreen}, D.~M. 2008, \apj, 688, 67,
  \dodoi{10.1086/592190}

\bibitem[{Elmegreen \& Elmegreen(2005)}]{Elmegreen_2005}
Elmegreen, B.~G., \& Elmegreen, D.~M. 2005, The Astrophysical Journal, 627,
  632, \dodoi{10.1086/430514}

\bibitem[{{Elmegreen} \& {Elmegreen}(2014)}]{Elmegreen_2014}
{Elmegreen}, D.~M., \& {Elmegreen}, B.~G. 2014, \apj, 781, 11,
  \dodoi{10.1088/0004-637X/781/1/11}

\bibitem[{{Elmegreen} {et~al.}(2009){Elmegreen}, {Elmegreen}, {Marcus},
  {Shahinyan}, {Yau}, \& {Petersen}}]{elmegreen_2009}
{Elmegreen}, D.~M., {Elmegreen}, B.~G., {Marcus}, M.~T., {et~al.} 2009, \apj,
  701, 306, \dodoi{10.1088/0004-637X/701/1/306}

\bibitem[{{Ferreira} {et~al.}(2022){Ferreira}, {Adams}, {Conselice},
  {Sazonova}, {Austin}, {Caruana}, {Ferrari}, {Verma}, {Trussler},
  {Broadhurst}, {Diego}, {Frye}, {Pascale}, {Wilkins}, {Windhorst}, \&
  {Zitrin}}]{Ferreira_2022}
{Ferreira}, L., {Adams}, N., {Conselice}, C.~J., {et~al.} 2022, \apjl, 938, L2,
  \dodoi{10.3847/2041-8213/ac947c}

\bibitem[{Ferreira {et~al.}(2023)Ferreira, Conselice, Sazonova, Ferrari,
  Caruana, Tohill, Lucatelli, Adams, Irodotou, Marshall, Roper, Lovell, Verma,
  Austin, Trussler, \& Wilkins}]{Ferreira_2023}
Ferreira, L., Conselice, C.~J., Sazonova, E., {et~al.} 2023, The Astrophysical
  Journal, 955, 94, \dodoi{10.3847/1538-4357/acec76}

\bibitem[{Finkelstein {et~al.}(2023)Finkelstein, Bagley, Ferguson, Wilkins,
  Kartaltepe, Papovich, Yung, Haro, Behroozi, Dickinson, Kocevski, Koekemoer,
  Larson, Bail, Morales, Pérez-González, Burgarella, Davé, Hirschmann,
  Somerville, Wuyts, Bromm, Casey, Fontana, Fujimoto, Gardner, Giavalisco,
  Grazian, Grogin, Hathi, Hutchison, Jha, Jogee, Kewley, Kirkpatrick, Long,
  Lotz, Pentericci, Pierel, Pirzkal, Ravindranath, Ryan, Trump, Yang,
  Bhatawdekar, Bisigello, Buat, Calabrò, Castellano, Cleri, Cooper, Croton,
  Daddi, Dekel, Elbaz, Franco, Gawiser, Holwerda, Huertas-Company, Jaskot,
  Leung, Lucas, Mobasher, Pandya, Tacchella, Weiner, \&
  Zavala}]{Finkelstein_2023}
Finkelstein, S.~L., Bagley, M.~B., Ferguson, H.~C., {et~al.} 2023, The
  Astrophysical Journal Letters, 946, L13, \dodoi{10.3847/2041-8213/acade4}

\bibitem[{{Fraternali} {et~al.}(2021){Fraternali}, {Karim}, {Magnelli},
  {G{\'o}mez-Guijarro}, {Jim{\'e}nez-Andrade}, \& {Posses}}]{fraternali_2021}
{Fraternali}, F., {Karim}, A., {Magnelli}, B., {et~al.} 2021, \aap, 647, A194,
  \dodoi{10.1051/0004-6361/202039807}

\bibitem[{{Genzel} {et~al.}(2011){Genzel}, {Newman}, {Jones}, {F{\"o}rster
  Schreiber}, {Shapiro}, {Genel}, {Lilly}, {Renzini}, {Tacconi}, {Bouch{\'e}},
  {Burkert}, {Cresci}, {Buschkamp}, {Carollo}, {Ceverino}, {Davies}, {Dekel},
  {Eisenhauer}, {Hicks}, {Kurk}, {Lutz}, {Mancini}, {Naab}, {Peng},
  {Sternberg}, {Vergani}, \& {Zamorani}}]{genzel_2011}
{Genzel}, R., {Newman}, S., {Jones}, T., {et~al.} 2011, \apj, 733, 101,
  \dodoi{10.1088/0004-637X/733/2/101}

\bibitem[{{Giavalisco} {et~al.}(1996){Giavalisco}, {Livio}, {Bohlin},
  {Macchetto}, \& {Stecher}}]{giavalisco_1996}
{Giavalisco}, M., {Livio}, M., {Bohlin}, R.~C., {Macchetto}, F.~D., \&
  {Stecher}, T.~P. 1996, \aj, 112, 369, \dodoi{10.1086/118021}

\bibitem[{{Gittins} \& {Clarke}(2004)}]{gittins_2004}
{Gittins}, D.~M., \& {Clarke}, C.~J. 2004, \mnras, 349, 909,
  \dodoi{10.1111/j.1365-2966.2004.07560.x}

\bibitem[{{Goldreich} \& {Lynden-Bell}(1965)}]{goldreich_1965}
{Goldreich}, P., \& {Lynden-Bell}, D. 1965, \mnras, 130, 125,
  \dodoi{10.1093/mnras/130.2.125}

\bibitem[{{Grogin} {et~al.}(2011){Grogin}, {Kocevski}, {Faber}, {Ferguson},
  {Koekemoer}, {Riess}, {Acquaviva}, {Alexander}, {Almaini}, {Ashby}, {Barden},
  {Bell}, {Bournaud}, {Brown}, {Caputi}, {Casertano}, {Cassata}, {Castellano},
  {Challis}, {Chary}, {Cheung}, {Cirasuolo}, {Conselice}, {Roshan Cooray},
  {Croton}, {Daddi}, {Dahlen}, {Dav{\'e}}, {de Mello}, {Dekel}, {Dickinson},
  {Dolch}, {Donley}, {Dunlop}, {Dutton}, {Elbaz}, {Fazio}, {Filippenko},
  {Finkelstein}, {Fontana}, {Gardner}, {Garnavich}, {Gawiser}, {Giavalisco},
  {Grazian}, {Guo}, {Hathi}, {H{\"a}ussler}, {Hopkins}, {Huang}, {Huang},
  {Jha}, {Kartaltepe}, {Kirshner}, {Koo}, {Lai}, {Lee}, {Li}, {Lotz}, {Lucas},
  {Madau}, {McCarthy}, {McGrath}, {McIntosh}, {McLure}, {Mobasher},
  {Moustakas}, {Mozena}, {Nandra}, {Newman}, {Niemi}, {Noeske}, {Papovich},
  {Pentericci}, {Pope}, {Primack}, {Rajan}, {Ravindranath}, {Reddy}, {Renzini},
  {Rix}, {Robaina}, {Rodney}, {Rosario}, {Rosati}, {Salimbeni}, {Scarlata},
  {Siana}, {Simard}, {Smidt}, {Somerville}, {Spinrad}, {Straughn}, {Strolger},
  {Telford}, {Teplitz}, {Trump}, {van der Wel}, {Villforth}, {Wechsler},
  {Weiner}, {Wiklind}, {Wild}, {Wilson}, {Wuyts}, {Yan}, \&
  {Yun}}]{grogin_2011}
{Grogin}, N.~A., {Kocevski}, D.~D., {Faber}, S.~M., {et~al.} 2011, \apjs, 197,
  35, \dodoi{10.1088/0067-0049/197/2/35}

\bibitem[{{Grosb{\o}l} \& {Dottori}(2012)}]{grosbol/dottori_2012}
{Grosb{\o}l}, P., \& {Dottori}, H. 2012, \aap, 542, A39,
  \dodoi{10.1051/0004-6361/201118099}

\bibitem[{{Grosbol} \& {Patsis}(1998)}]{grosbol_1998}
{Grosbol}, P.~J., \& {Patsis}, P.~A. 1998, \aap, 336, 840

\bibitem[{{Guo} {et~al.}(2012){Guo}, {Giavalisco}, {Ferguson}, {Cassata}, \&
  {Koekemoer}}]{guo_2012}
{Guo}, Y., {Giavalisco}, M., {Ferguson}, H.~C., {Cassata}, P., \& {Koekemoer},
  A.~M. 2012, \apj, 757, 120, \dodoi{10.1088/0004-637X/757/2/120}

\bibitem[{{Guo} {et~al.}(2015){Guo}, {Ferguson}, {Bell}, {Koo}, {Conselice},
  {Giavalisco}, {Kassin}, {Lu}, {Lucas}, {Mandelker}, {McIntosh}, {Primack},
  {Ravindranath}, {Barro}, {Ceverino}, {Dekel}, {Faber}, {Fang}, {Koekemoer},
  {Noeske}, {Rafelski}, \& {Straughn}}]{guo_2015}
{Guo}, Y., {Ferguson}, H.~C., {Bell}, E.~F., {et~al.} 2015, \apj, 800, 39,
  \dodoi{10.1088/0004-637X/800/1/39}

\bibitem[{{Guo} {et~al.}(2018){Guo}, {Rafelski}, {Bell}, {Conselice}, {Dekel},
  {Faber}, {Giavalisco}, {Koekemoer}, {Koo}, {Lu}, {Mandelker}, {Primack},
  {Ceverino}, {de Mello}, {Ferguson}, {Hathi}, {Kocevski}, {Lucas},
  {P{\'e}rez-Gonz{\'a}lez}, {Ravindranath}, {Soto}, {Straughn}, \&
  {Wang}}]{guo_2018}
{Guo}, Y., {Rafelski}, M., {Bell}, E.~F., {et~al.} 2018, \apj, 853, 108,
  \dodoi{10.3847/1538-4357/aaa018}

\bibitem[{Hart {et~al.}(2017)Hart, Bamford, Casteels, Kruk, Lintott, \&
  Masters}]{Hart_2017}
Hart, R.~E., Bamford, S.~P., Casteels, K. R.~V., {et~al.} 2017, Monthly Notices
  of the Royal Astronomical Society, 468, 1850, \dodoi{10.1093/mnras/stx581}

\bibitem[{Huang {et~al.}(2023)Huang, Kawabe, Kohno, Saito, Mizukoshi, Iono,
  Michiyama, Tamura, Hayward, \& Umehata}]{Huang_2023}
Huang, S., Kawabe, R., Kohno, K., {et~al.} 2023, The Astrophysical Journal
  Letters, 958, L26, \dodoi{10.3847/2041-8213/acff63}

\bibitem[{{Julian} \& {Toomre}(1966)}]{julian/toomre_1966}
{Julian}, W.~H., \& {Toomre}, A. 1966, \apj, 146, 810, \dodoi{10.1086/148957}

\bibitem[{{Kartaltepe} {et~al.}(2023){Kartaltepe}, {Rose}, {Vanderhoof},
  {McGrath}, {Costantin}, {Cox}, {Yung}, {Kocevski}, {Wuyts}, {Ferguson},
  {Bagley}, {Finkelstein}, {Amor{\'\i}n}, {Andrews}, {Haro}, {Backhaus},
  {Behroozi}, {Bisigello}, {Calabr{\`o}}, {Casey}, {Coogan}, {Cooper},
  {Croton}, {de la Vega}, {Dickinson}, {Fontana}, {Franco}, {Grazian},
  {Grogin}, {Hathi}, {Holwerda}, {Huertas-Company}, {Iyer}, {Jogee}, {Jung},
  {Kewley}, {Kirkpatrick}, {Koekemoer}, {Liu}, {Lotz}, {Lucas}, {Newman},
  {Pacifici}, {Pandya}, {Papovich}, {Pentericci}, {P{\'e}rez-Gonz{\'a}lez},
  {Petersen}, {Pirzkal}, {Rafelski}, {Ravindranath}, {Simons}, {Snyder},
  {Somerville}, {Stanway}, {Straughn}, {Tacchella}, {Trump}, {Vega-Ferrero},
  {Wilkins}, {Yang}, \& {Zavala}}]{Kartaltepe_2023}
{Kartaltepe}, J.~S., {Rose}, C., {Vanderhoof}, B.~N., {et~al.} 2023, \apjl,
  946, L15, \dodoi{10.3847/2041-8213/acad01}

\bibitem[{{Kendall} {et~al.}(2015){Kendall}, {Clarke}, \&
  {Kennicutt}}]{kendall_2015}
{Kendall}, S., {Clarke}, C., \& {Kennicutt}, R.~C. 2015, \mnras, 446, 4155,
  \dodoi{10.1093/mnras/stu2431}

\bibitem[{Kennicutt(1998)}]{Kennicutt_1998}
Kennicutt, R.~C. 1998, Annual Review of Astronomy and Astrophysics, 36, 189,
  \dodoi{10.1146/annurev.astro.36.1.189}

\bibitem[{{Kim} {et~al.}(2010){Kim}, {Kim}, \& {Ostriker}}]{kim_2010}
{Kim}, C.-G., {Kim}, W.-T., \& {Ostriker}, E.~C. 2010, \apj, 720, 1454,
  \dodoi{10.1088/0004-637X/720/2/1454}

\bibitem[{{Koekemoer} {et~al.}(2011){Koekemoer}, {Faber}, {Ferguson}, {Grogin},
  {Kocevski}, {Koo}, {Lai}, {Lotz}, {Lucas}, {McGrath}, {Ogaz}, {Rajan},
  {Riess}, {Rodney}, {Strolger}, {Casertano}, {Castellano}, {Dahlen},
  {Dickinson}, {Dolch}, {Fontana}, {Giavalisco}, {Grazian}, {Guo}, {Hathi},
  {Huang}, {van der Wel}, {Yan}, {Acquaviva}, {Alexander}, {Almaini}, {Ashby},
  {Barden}, {Bell}, {Bournaud}, {Brown}, {Caputi}, {Cassata}, {Challis},
  {Chary}, {Cheung}, {Cirasuolo}, {Conselice}, {Roshan Cooray}, {Croton},
  {Daddi}, {Dav{\'e}}, {de Mello}, {de Ravel}, {Dekel}, {Donley}, {Dunlop},
  {Dutton}, {Elbaz}, {Fazio}, {Filippenko}, {Finkelstein}, {Frazer}, {Gardner},
  {Garnavich}, {Gawiser}, {Gruetzbauch}, {Hartley}, {H{\"a}ussler},
  {Herrington}, {Hopkins}, {Huang}, {Jha}, {Johnson}, {Kartaltepe},
  {Khostovan}, {Kirshner}, {Lani}, {Lee}, {Li}, {Madau}, {McCarthy},
  {McIntosh}, {McLure}, {McPartland}, {Mobasher}, {Moreira}, {Mortlock},
  {Moustakas}, {Mozena}, {Nandra}, {Newman}, {Nielsen}, {Niemi}, {Noeske},
  {Papovich}, {Pentericci}, {Pope}, {Primack}, {Ravindranath}, {Reddy},
  {Renzini}, {Rix}, {Robaina}, {Rosario}, {Rosati}, {Salimbeni}, {Scarlata},
  {Siana}, {Simard}, {Smidt}, {Snyder}, {Somerville}, {Spinrad}, {Straughn},
  {Telford}, {Teplitz}, {Trump}, {Vargas}, {Villforth}, {Wagner}, {Wandro},
  {Wechsler}, {Weiner}, {Wiklind}, {Wild}, {Wilson}, {Wuyts}, \&
  {Yun}}]{koekemoer_2011}
{Koekemoer}, A.~M., {Faber}, S.~M., {Ferguson}, H.~C., {et~al.} 2011, \apjs,
  197, 36, \dodoi{10.1088/0067-0049/197/2/36}

\bibitem[{{Kormendy} \& {Norman}(1979)}]{kormendy_1979}
{Kormendy}, J., \& {Norman}, C.~A. 1979, \apj, 233, 539, \dodoi{10.1086/157414}

\bibitem[{Law {et~al.}(2012)Law, Shapley, Steidel, Reddy, Christensen, \&
  Erb}]{Law_2012}
Law, D.~R., Shapley, A.~E., Steidel, C.~C., {et~al.} 2012, Nature, 487, 338,
  \dodoi{10.1038/nature11256}

\bibitem[{{Lelli} {et~al.}(2018){Lelli}, {De Breuck}, {Falkendal},
  {Fraternali}, {Man}, {Nesvadba}, \& {Lehnert}}]{lelli_2018}
{Lelli}, F., {De Breuck}, C., {Falkendal}, T., {et~al.} 2018, \mnras, 479,
  5440, \dodoi{10.1093/mnras/sty1795}

\bibitem[{{Lelli} {et~al.}(2021){Lelli}, {Di Teodoro}, {Fraternali}, {Man},
  {Zhang}, {De Breuck}, {Davis}, \& {Maiolino}}]{lelli_2021}
{Lelli}, F., {Di Teodoro}, E.~M., {Fraternali}, F., {et~al.} 2021, Science,
  371, 713, \dodoi{10.1126/science.abc1893}

\bibitem[{{Lin} \& {Shu}(1964)}]{Lin_1964}
{Lin}, C.~C., \& {Shu}, F.~H. 1964, \apj, 140, 646, \dodoi{10.1086/147955}

\bibitem[{Margalef-Bentabol {et~al.}(2022)Margalef-Bentabol, Conselice,
  Haeussler, Casteels, Lintott, Masters, \& Simmons}]{Margalef-Bentabol_2022}
Margalef-Bentabol, B., Conselice, C.~J., Haeussler, B., {et~al.} 2022, Monthly
  Notices of the Royal Astronomical Society, 511, 1502,
  \dodoi{10.1093/mnras/stac080}

\bibitem[{{Margalef-Bentabol} {et~al.}(2016){Margalef-Bentabol}, {Conselice},
  {Mortlock}, {Hartley}, {Duncan}, {Ferguson}, {Dekel}, \&
  {Primack}}]{margalef-bentabol_2016}
{Margalef-Bentabol}, B., {Conselice}, C.~J., {Mortlock}, A., {et~al.} 2016,
  \mnras, 461, 2728, \dodoi{10.1093/mnras/stw1451}

\bibitem[{{Martin} {et~al.}(2023){Martin}, {Guo}, {Wang}, {Koekemoer},
  {Rafelski}, {Teplitz}, {Windhorst}, {Alavi}, {Grogin}, {Prichard},
  {Sunnquist}, {Ceverino}, {Chartab}, {Conselice}, {Dai}, {Dekel}, {Gardner},
  {Gawiser}, {Hathi}, {Hayes}, {Jansen}, {Ji}, {Koo}, {Lucas}, {Mandelker},
  {Mehta}, {Mobasher}, {Nedkova}, {Primack}, {Ravindranath}, {Robertson},
  {Rutkowski}, {Sattari}, {Soto}, \& {Yung}}]{martin_2023}
{Martin}, A., {Guo}, Y., {Wang}, X., {et~al.} 2023, \apj, 955, 106,
  \dodoi{10.3847/1538-4357/aced3e}

\bibitem[{{Mobasher} {et~al.}(2015){Mobasher}, {Dahlen}, {Ferguson},
  {Acquaviva}, {Barro}, {Finkelstein}, {Fontana}, {Gruetzbauch}, {Johnson},
  {Lu}, {Papovich}, {Pforr}, {Salvato}, {Somerville}, {Wiklind}, {Wuyts},
  {Ashby}, {Bell}, {Conselice}, {Dickinson}, {Faber}, {Fazio}, {Finlator},
  {Galametz}, {Gawiser}, {Giavalisco}, {Grazian}, {Grogin}, {Guo}, {Hathi},
  {Kocevski}, {Koekemoer}, {Koo}, {Newman}, {Reddy}, {Santini}, \&
  {Wechsler}}]{mobasher_2015}
{Mobasher}, B., {Dahlen}, T., {Ferguson}, H.~C., {et~al.} 2015, \apj, 808, 101,
  \dodoi{10.1088/0004-637X/808/1/101}

\bibitem[{{Nelson} {et~al.}(2023){Nelson}, {Brammer}, {Gimenez-Arteaga},
  {Oesch}, {Ubler}, {de Graaff}, {Matharu}, {Naidu}, {Shapley}, {Whitaker},
  {Wisnioski}, {Forster Schreiber}, {Smit}, {van Dokkum}, {Chisholm},
  {Endsley}, {Hartley}, {Gibson}, {Giovinazzo}, {Illingworth}, {Labbe},
  {Maseda}, {Matthee}, {Covelo Paz}, {Price}, {Reddy}, {Shivaei}, {Weibel},
  {Wuyts}, {Xiao}, {Alberts}, {Baker}, {Bunker}, {Cameron}, {Charlot},
  {Eisenstein}, {Ji}, {Johnson}, {Jones}, {Maiolino}, {Robertson}, {Sandles},
  {Suess}, {Tacchella}, {Williams}, \& {Witstok}}]{nelson_2023}
{Nelson}, E.~J., {Brammer}, G., {Gimenez-Arteaga}, C., {et~al.} 2023, arXiv
  e-prints, arXiv:2310.06887, \dodoi{10.48550/arXiv.2310.06887}

\bibitem[{Porter-Temple {et~al.}(2022)Porter-Temple, Holwerda, Hopkins, Porter,
  Henry, Geron, Simmons, Masters, \& Kruk}]{Porter_Temple_2022}
Porter-Temple, R., Holwerda, B.~W., Hopkins, A.~M., {et~al.} 2022, Monthly
  Notices of the Royal Astronomical Society, 515, 3875,
  \dodoi{10.1093/mnras/stac1936}

\bibitem[{{Rizzo} {et~al.}(2021){Rizzo}, {Vegetti}, {Fraternali}, {Stacey}, \&
  {Powell}}]{rizzo_2021}
{Rizzo}, F., {Vegetti}, S., {Fraternali}, F., {Stacey}, H.~R., \& {Powell}, D.
  2021, \mnras, 507, 3952, \dodoi{10.1093/mnras/stab2295}

\bibitem[{{Rizzo} {et~al.}(2020){Rizzo}, {Vegetti}, {Powell}, {Fraternali},
  {McKean}, {Stacey}, \& {White}}]{rizzo_2020}
{Rizzo}, F., {Vegetti}, S., {Powell}, D., {et~al.} 2020, \nat, 584, 201,
  \dodoi{10.1038/s41586-020-2572-6}

\bibitem[{Sellwood \& Masters(2022)}]{sellwood_2022}
Sellwood, J., \& Masters, K.~L. 2022, Annual Review of Astronomy and
  Astrophysics, 60, 73, \dodoi{10.1146/annurev-astro-052920-104505}

\bibitem[{{Simons} {et~al.}(2017){Simons}, {Kassin}, {Weiner}, {Faber},
  {Trump}, {Heckman}, {Koo}, {Pacifici}, {Primack}, {Snyder}, \& {de la
  Vega}}]{simons_2017}
{Simons}, R.~C., {Kassin}, S.~A., {Weiner}, B.~J., {et~al.} 2017, \apj, 843,
  46, \dodoi{10.3847/1538-4357/aa740c}

\bibitem[{Stefanon {et~al.}(2017)Stefanon, Yan, Mobasher, Barro, Donley,
  Fontana, Hemmati, Koekemoer, Lee, Lee, Nayyeri, Peth, Pforr, Salvato,
  Wiklind, Wuyts, Ashby, Castellano, Conselice, Cooper, Cooray, Dolch,
  Ferguson, Galametz, Giavalisco, Guo, Willner, Dickinson, Faber, Fazio,
  Gardner, Gawiser, Grazian, Grogin, Kocevski, Koo, Lee, Lucas, McGrath,
  Nandra, Newman, \& van~der Wel}]{Stefanon_2017}
Stefanon, M., Yan, H., Mobasher, B., {et~al.} 2017, The Astrophysical Journal
  Supplement Series, 229, 32, \dodoi{10.3847/1538-4365/aa66cb}

\bibitem[{{Toomre}(1964)}]{toomre_1964}
{Toomre}, A. 1964, \apj, 139, 1217, \dodoi{10.1086/147861}

\bibitem[{{Tsukui} \& {Iguchi}(2021)}]{tsukui_2021}
{Tsukui}, T., \& {Iguchi}, S. 2021, Science, 372, 1201,
  \dodoi{10.1126/science.abe9680}

\bibitem[{{van der Wel} {et~al.}(2012){van der Wel}, {Bell}, {H{\"a}ussler},
  {McGrath}, {Chang}, {Guo}, {McIntosh}, {Rix}, {Barden}, {Cheung}, {Faber},
  {Ferguson}, {Galametz}, {Grogin}, {Hartley}, {Kartaltepe}, {Kocevski},
  {Koekemoer}, {Lotz}, {Mozena}, {Peth}, \& {Peng}}]{vanderwel_2012}
{van der Wel}, A., {Bell}, E.~F., {H{\"a}ussler}, B., {et~al.} 2012, \apjs,
  203, 24, \dodoi{10.1088/0067-0049/203/2/24}

\bibitem[{{van der Wel} {et~al.}(2014){van der Wel}, {Franx}, {van Dokkum},
  {Skelton}, {Momcheva}, {Whitaker}, {Brammer}, {Bell}, {Rix}, {Wuyts},
  {Ferguson}, {Holden}, {Barro}, {Koekemoer}, {Chang}, {McGrath},
  {H{\"a}ussler}, {Dekel}, {Behroozi}, {Fumagalli}, {Leja}, {Lundgren},
  {Maseda}, {Nelson}, {Wake}, {Patel}, {Labb{\'e}}, {Faber}, {Grogin}, \&
  {Kocevski}}]{van_der_wel_2014}
{van der Wel}, A., {Franx}, M., {van Dokkum}, P.~G., {et~al.} 2014, \apj, 788,
  28, \dodoi{10.1088/0004-637X/788/1/28}

\bibitem[{{Wen} \& {Zheng}(2016)}]{wen_2016}
{Wen}, Z.~Z., \& {Zheng}, X.~Z. 2016, \apj, 832, 90,
  \dodoi{10.3847/0004-637X/832/1/90}

\bibitem[{{Whitaker} {et~al.}(2014){Whitaker}, {Franx}, {Leja}, {van Dokkum},
  {Henry}, {Skelton}, {Fumagalli}, {Momcheva}, {Brammer}, {Labb{\'e}},
  {Nelson}, \& {Rigby}}]{whitaker_2014}
{Whitaker}, K.~E., {Franx}, M., {Leja}, J., {et~al.} 2014, \apj, 795, 104,
  \dodoi{10.1088/0004-637X/795/2/104}

\bibitem[{{Wu} {et~al.}(2023){Wu}, {Cai}, {Sun}, {Bian}, {Lin}, {Li}, {Li},
  {Bauer}, {Egami}, {Fan}, {Gonz{\'a}lez-L{\'o}pez}, {Li}, {Wang}, {Yang},
  {Zhang}, \& {Zou}}]{Wu_2023}
{Wu}, Y., {Cai}, Z., {Sun}, F., {et~al.} 2023, \apjl, 942, L1,
  \dodoi{10.3847/2041-8213/aca652}

\bibitem[{{Yu} {et~al.}(2023){Yu}, {Cheng}, {Pan}, {Sun}, \& {Li}}]{yu_2023}
{Yu}, S.-Y., {Cheng}, C., {Pan}, Y., {Sun}, F., \& {Li}, Y.~A. 2023, \aap, 676,
  A74, \dodoi{10.1051/0004-6361/202346140}

\bibitem[{{Yu} \& {Ho}(2018)}]{yu_2018}
{Yu}, S.-Y., \& {Ho}, L.~C. 2018, \apj, 869, 29,
  \dodoi{10.3847/1538-4357/aaeacd}

\bibitem[{{Yu} {et~al.}(2021){Yu}, {Ho}, \& {Wang}}]{yu_2021}
{Yu}, S.-Y., {Ho}, L.~C., \& {Wang}, J. 2021, \apj, 917, 88,
  \dodoi{10.3847/1538-4357/ac0c77}

\end{thebibliography}
\bibliographystyle{aasjournal}

\end{document}